\documentclass[aps,prb,twocolumn,preprintnumbers,superscriptaddress,amsmath]{revtex4}

\usepackage{graphicx}
\usepackage{dcolumn}
\usepackage{bm}
\usepackage{soul}
\usepackage{color}
\usepackage{xcolor}
\usepackage{amsmath}
\usepackage{amssymb}
\usepackage{amsmath}
\usepackage{float}
\usepackage{natmove}
\usepackage{multirow}
\usepackage{setspace}
\usepackage{array}
\usepackage{booktabs}
\usepackage{rotating}
\usepackage{ulem}
\usepackage[colorlinks=true,linkcolor=blue,citecolor=blue,urlcolor=blue]{hyperref}
\usepackage{physics}
\usepackage{mathrsfs}
\definecolor{Grey}{rgb}{0.50,0.50,0.50}
\definecolor{Blu}{rgb}{0.00,0.00,1.00}
\definecolor{Red}{rgb}{1.00,0.00,0.00}
\definecolor{Green}{rgb}{0.00,0.60,0.00}
\definecolor{Magenta}{rgb}{0.60,0.00,0.60}
\definecolor{BluBondi}{rgb}{0.00,0.58,0.71}
\definecolor{Orange}{rgb}{0.95,0.46,0.17}

\newcommand{\editor}[2]{%
  \expandafter\newcommand\csname #1note\endcsname[1]{%
    \textcolor{#2}{(\textbf{#1:} \it ##1)}}%
  \expandafter\newcommand\csname #1\endcsname[1]{%
    \textcolor{#2}{##1}}%
  \expandafter\newcommand\csname #1cancel\endcsname[1]{%
    \textcolor{#2}{\sout{##1}}}%
  \expandafter\newcommand\csname #1change\endcsname[2]{%
    \textcolor{#2}{\sout{##1} ##2}}%
  \newenvironment{#1text}{\color{#2}}{\color{black}}
}

\editor{MB}{orange}
\editor{EM}{BluBondi}
\editor{MZ}{magenta}
\editor{AF}{BluBondi}
\editor{DV}{Blu}
\editor{AR}{red}
\editor{MJC}{green}

\newcommand{\suppinfo}{Supplemental Material~\cite{supp-info}}
\newcommand{\drop}[1]{}

\begin{document}
\title{Quenching of low-energy optical absorption in bilayer C$_3$N polytypes}
\author{Matteo Zanfrognini}
\email[corresponding author: ]{matteo.zanfrognini@unimore.it}
\affiliation{Dipartimento di Scienze Fisiche, Informatiche e Matematiche, Universit$\grave{a}$ di Modena e Reggio Emilia, I-41125 Modena, Italy}
\affiliation{Centro S3, CNR-Istituto Nanoscienze, I-41125 Modena, Italy}
\author{Miki Bonacci}
\affiliation{Dipartimento di Scienze Fisiche, Informatiche e Matematiche, Universit$\grave{a}$ di Modena e Reggio Emilia, I-41125 Modena, Italy}
\affiliation{Centro S3, CNR-Istituto Nanoscienze, I-41125 Modena, Italy}
\author{Fulvio Paleari}
\affiliation{Centro S3, CNR-Istituto Nanoscienze, I-41125 Modena, Italy}
\author{Elisa Molinari}
\affiliation{Dipartimento di Scienze Fisiche, Informatiche e Matematiche, Universit$\grave{a}$ di Modena e Reggio Emilia, I-41125 Modena, Italy}
\affiliation{Centro S3, CNR-Istituto Nanoscienze, I-41125 Modena, Italy}
\author{Alice Ruini}
\affiliation{Dipartimento di Scienze Fisiche, Informatiche e Matematiche, Universit$\grave{a}$ di Modena e Reggio Emilia, I-41125 Modena, Italy}
\affiliation{Centro S3, CNR-Istituto Nanoscienze, I-41125 Modena, Italy}
\author{Andrea Ferretti}
\affiliation{Centro S3, CNR-Istituto Nanoscienze, I-41125 Modena, Italy}
\author{Marilia J. Caldas}
\affiliation{Instituto de F\'\i{}sica, Universidade de S{\~a}o Paulo,
  Cidade Universit\'aria, 05508-900 S{\~a}o Paulo, Brazil}
\author{Daniele Varsano}
\email{daniele.varsano@nano.cnr.it}
\affiliation{Centro S3, CNR-Istituto Nanoscienze, I-41125 Modena, Italy}
\date{\today}
\begin{abstract}
In this work we provide a first principles description of the electronic and optical properties of bilayers C$_3$N, with different stacking motifs AB, AB$'$ and AA$'$. Starting from quasi-particle electronic band-structures, we solve the Bethe Salpeter Equation (BSE) to access the excitonic properties of these bilayers. For all stacking sequences, we see strong optical absorption at energies lower than but close to that of the monolayer. Most relevant, we predict a strong quenching of the low-energy optical absorption, with  negligible oscillator strength of low-lying bound excitons. This is a unique phenomenology that does not arise in the monolayer case, 
nor in other common homo-bilayers. We explain these findings in terms of the small interband dipole matrix elements associated to the valence-conduction transitions involved in these excitons, and discuss them in view of the different stacking motifs.

\end{abstract}
\maketitle
%
\section{Introduction}
\label{sec:Introduction}
%
Many recent experimental and theoretical works have focused on a detailed understanding of the electronic and optical properties of two-dimensional (2D) materials, due to their potential use in the design of innovative optoelectronic devices which could combine atomic-size dimensions with improved performance~\cite{Mak_2010,Zhang_2015,Chernikov_2014,Cudazzo_2016,Lin_2016,Rousseau_2021,Zhang_2017,Galvani_2016,Tran_2014,Thygesen_2017,Palummo_2015}. 
Great attention has been reserved to the study of Van der Waals structures, where atomically-thin 2D materials are vertically stacked and held together by weak and long range dispersive interactions, strongly affecting both the optical~\cite{Andersen_2015,Paleari_2018} and the electronic properties of the isolated single layers~\cite{He_2014}.
A promising material for such novel applications is graphene-like 2D polyaniline (also known as monolayer C$_3$N), which has been recently synthesized through different approaches~\cite{Mahmood_2016,Yang_2017}: its electronic and optical properties have been extensively studied from a theoretical point of view~\cite{Wu_2018,Miki_2022,Zhao_2022,io_2023}, revealing a quasi-particle band structure with  indirect band-gap and  intense optical absorption for photons in a narrow spectral range around 2 eV.

Vertical stacking of two layers of C$_3$N is a possible approach to tune its electronic and optical properties.
Few theoretical works have analysed the stability of bilayers C$_3$N (BL-C$_3$N) as a function of the possible stacking patterns~\cite{Wu_2018,Wei_2021}. Following the notation of Ref.~[\onlinecite{Wei_2021}], all these calculations have obtained negative formation energies for  AB and AB$'$ (displaced-like), 
as well as AA$'$ (sandwich-like stacking) arrangements.
Furthermore, BL-C$_3$N with AA$'$ and AB$'$ stackings have also been experimentally synthesized, as described in Ref.~[\onlinecite{Wei_2021}], where a detailed investigation of the electronic properties using scanning tunnelling spectroscopy (STS) has been provided. The results indicate a strong change of the electronic-transport band gap passing from monolayer to bilayer, together with relevant modifications of the electronic properties as a function of the stacking sequence.

Motivated by these experimental advances, in this work we discuss the optical properties of BL-C$_3$N via first principles methods, properly including excitonic effects which are known to play a fundamental role in the description of optical absorption in 2D-materials~\cite{Latini_2015,Andersen_2015,Falco_2013}. As a first result, we find for all systems a strong optical absorption in an energy region around 1.7 {eV} which is sightly lower than the absorption peak of the isolated monolayer~\cite{Miki_2022}, despite the consistent reduction of the electronic gap. This feature, combined with the energy band structure associated to these absorption peaks, is promising for photovoltaics. 
Moreover, our results indicate that, for all the considered stackings, the optical response does not present other bright excitonic states at lower energies. 
Such behaviour is not observed in other common semiconducting bilayer homo-structures, e.g. BL-hBN\cite{Paleari_2018}, BL-MoS$_2$\cite{Mak_2010,Palummo_2015} or BL-phosphorene\cite{Tran_2014}.
Finally, the origin of this low energy absorption quenching is explained and rationalized, focusing on the properties of the single particle states involved in the formation of the lowest-energy excitons. 

The article is organized as follows. In Section \ref{sec:methods}, we summarize the computational methods used within this work, while in Section \ref{sec:structure_and_bands} we present quasi-particle band structures for the AB and AB$'$ 
displaced-like stacking patterns. In Section \ref{sec:optics} we discuss the optical properties of these bilayers and in Section \ref{sec:INTERBAND_DIPOLE} we provide a rationale for the negligible oscillator strength observed for the low-lying excitons. Finally, in Section \ref{sec:STACKING_AA1}, we discuss the optical properties of bilayer C$_3$N with AA$'$ 
sandwich-like stacking, predicting a strong optical absorption quenching starting from the symmetry properties of this polytype and confirming such interpretation with an approximate solution of the Bethe Salpeter equation. 

\section{Computational methods}
\label{sec:methods}
%
Ground state structural and electronic properties have been investigated using density functional theory (DFT), as implemented in the plane-wave-basis-set package {\sc{Quantum ESPRESSO}}~\cite{Giannozzi_2009,Giannozzi_2017}. In these calculations, we have used norm-conserving ONCV pseudopotentials~\cite{Hamann_2013}, within the GGA-PBE approximation~\cite{Perdew_1996} for the exchange-correlation potential. Van der Waals interactions between layers have been taken into account by adding the dispersion correction proposed by Grimme~\cite{Grimme_2006} to the exchange-correlation energy (PBE-D2).
Equilibrium structural properties have been obtained by relaxing both the in-plane unit cell and the atomic positions up to when the components of the forces acting on each atom were smaller than 5$\cdot$10$^{-4}$ Ry/Bohr. In all ground state calculations we  used a 12$\times$12$\times$1 Monkhorst-Pack~\cite{Monkhorst_1976} \textbf{k}-grid to sample the Brillouin Zone (BZ) and a kinetic energy cutoff of 90 Ry for the plane wave basis set used to represent single particle wavefunctions.

In the case of AB and AB$'$ stacking motifs, Kohn-Sham wavefunctions and eigenvalues, computed from the equilibrium ground state charge density, have been used to evaluate quasi-particle (QP) corrections to DFT energies within the GW approximation~\cite{Hedin_1965,Onida_2002} for the electron self energy. QP corrections $\epsilon_{n\mathbf{k}}^{\text{KS}}$ have been computed using the single shot G$_0$W$_0$ approach, evaluating the expectation value of the operator ${\Sigma} -{v}_{\text{KS}}$ on the Kohn-Sham states $|\psi_{n\mathbf{k}}\rangle$, being ${\Sigma}$ the electron self-energy operator and ${v}_{KS}$ the exchange-correlation potential. When solding the Dyson equation we linearized the frequency dependence of the self-energy, ${\Sigma}(E^{\text{QP}}_{n\mathbf{k}})\approx{\Sigma}(\epsilon^{\text{KS}}_{n\mathbf{k}}) + \frac{\partial{\Sigma}}{\partial\epsilon}\big|_{\epsilon^{\text{KS}}_{n\mathbf{k}}}(E^{\text{QP}}_{n\mathbf{k}} -\epsilon^{\text{KS}}_{n\mathbf{k}}) $, as proposed in Ref.~[\onlinecite{Hybertsen_1986}].

The electron-electron screened interaction $W$ has been computed using the Random Phase Approximation (RPA), as implemented in the {\tt{Yambo}} code~\cite{Marini_2009,Sangalli_2019}. Converged QP gaps within a 10 meV threshold required the inclusion of 700 bands and a $\mathbf{G}$-vector cutoff of 16 Ry in the construction of the screening matrix at the RPA level. The frequency dependence of $W$ has been described using the Godby-Needs plasmon-pole model~\cite{Godby_1989}, and 1000 bands have been included in the sum-over-states appearing in the correlation part of the  self energy. To reduce spurious interactions among different cells in the stacking direction, we have used a supercell length along $z$ of 23.5 \AA, together with a 2D cutoff~\cite{Beigi_2006,Rozzi_2006} for the Coulomb potential. Finally, to speed-up the convergence of QP gaps w.r.t. the \textbf{k}-point mesh, we have adopted the approach recently proposed by Guandalini~\textit{et al.}~\cite{Alberto_2022}.
In this work we have verified that, with this method, a 18$\times$18$\times$1 Monkhorst-Pack \textbf{k}-grid already provides converged gaps within the chosen threshold of 10 meV.

Starting from QP corrected electronic energies, we have obtained excitonic properties by solving the Bethe-Salpeter equation (BSE)~\cite{Strinati_1988,Rohlfing_2000} in the resonant (Tamm-Dancoff) approximation, i.e. via diagonalization of the excitonic Hamiltonian
\begin{equation}
    \begin{split}
    H_{exc}\big(vc\mathbf{k},v'c'\mathbf{k'}\big) &= \big (E_{c\mathbf{k}}^{\text{QP}} - E_{v\mathbf{k}}^{\text{QP}}\big)\delta_{cc'}\delta_{vv'} \\
    &+ K^{d}\big(vc\mathbf{k},v'c'\mathbf{k}'\big) + K^{x}\big(vc\mathbf{k},v'c'\mathbf{k}'\big)
    \end{split}
    \label{eq:Exc_H}
\end{equation}
where $K^{d}$ ($K^{x}$) is the direct (exchange) part of the BSE kernel, $(v,v')$ and $(c,c')$ are the valence and conduction bands included in the BSE and $E_{(v,c)\mathbf{k}}^{QP}$ are the QP corrected electron energies. Converged exciton energies have been obtained including the two highest-occupied valence and the four lowest unoccupied conduction bands in the construction of $H_{exc}\big(vc\mathbf{k},v'c'\mathbf{k'}\big)$,  and using a 48$\times$48$\times$1 Monkhorst-Pack \textbf{k}-grid to sample the BZ. We point out that, in the solution of the BSE, QP corrections have been approximated via a scissor-stretching operator (see \suppinfo{} for a detailed description about the fitting procedure), while the electron-electron screened interaction has been computed at the RPA level, in the static approximation, using the same converged parameters adopted for the calculation of QP corrections.

Finally, by diagonalization of $H_{exc}\big(vc\mathbf{k},v'c'\mathbf{k'}\big)$
\begin{equation}
    \sum_{v'c'\mathbf{k}'}H_{exc}\big(vc\mathbf{k},v'c'\mathbf{k'}\big)A_{\lambda}\big(v'c'\mathbf{k}'\big) = E_{\lambda}A_{\lambda}\big(vc\mathbf{k}\big)
    \label{eq:eq_exc_wfc}
\end{equation}
we obtained the energies $E_{\lambda}$ and the envelope functions $A_{\lambda}\big(vc\mathbf{k}\big)$ for each exciton $\lambda$. Optical absorption is finally computed as the imaginary part of the macroscopic dielectric function,
\begin{equation}
    \epsilon_{M}(E) = 1 - \frac{8\pi}{V}\sum_{\lambda}\frac{D_\lambda}{E - E_\lambda +i\eta}
    \label{eq:macro_eps}
\end{equation}
In the above expression, $V$ is the unit cell volume and $D_\lambda$ the oscillator strength of exciton $\lambda$ defined as
\begin{equation}
    D_\lambda = \bigg | \hat{\epsilon} \cdot \sum_{vc\mathbf{k}}d_{vc\mathbf{k}} A_{\lambda}(vc\mathbf{k})\bigg |^2,
    \label{eq:eq_OS}
\end{equation}
where $\hat{\epsilon}$ is the in-plane polarization direction of the incoming light, and $d_{vc\mathbf{k}} = \langle \varphi_{v\mathbf{k}}|\mathbf{r}|\varphi_{c\mathbf{k}}\rangle$ the single-particle interband dipole matrix element.
%
%
\section{Structural and electronic properties of bilayer C$_3$N with AB and AB$'$ stackings}
\label{sec:structure_and_bands}
%
%
\begin{figure}
  \centering
    \includegraphics[width=0.45\textwidth]{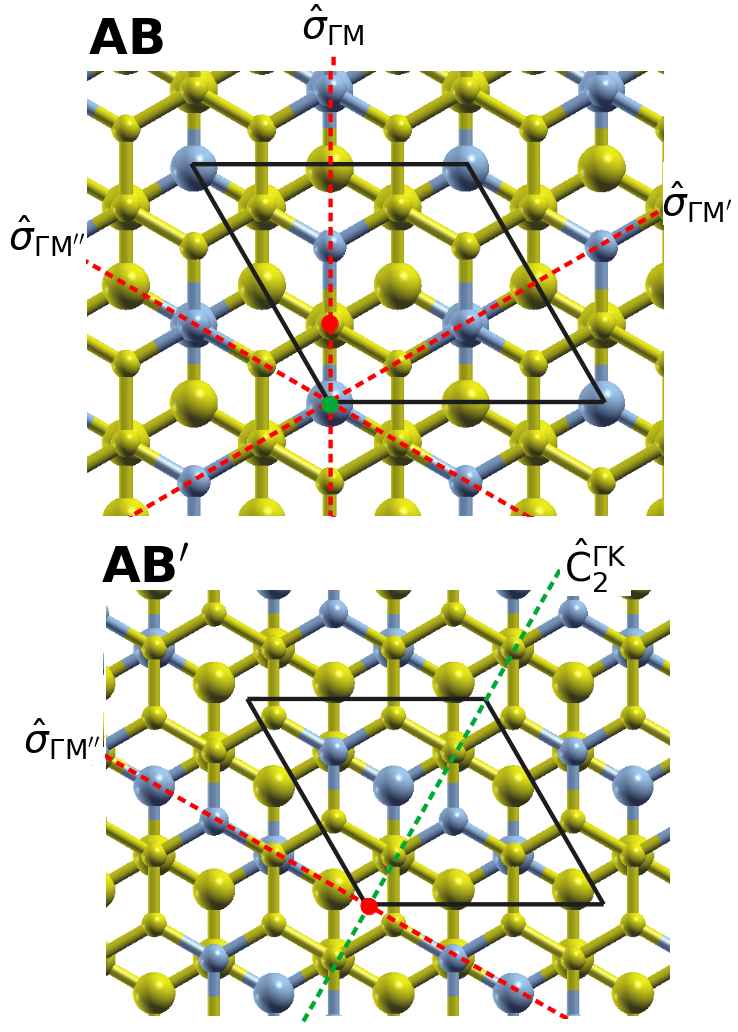}
    \caption{Crystal structures for bilayer C$_3$N with AB (upper panel) and AB$'$ (lower panel) stackings. Yellow (light blue) spheres indicate Carbon (Nitrogen) atoms, while small (large)  radius spheres denote atoms located on the upper (lower) layer. The red dot indicates the in-plane position of the inversion symmetry center, while dashed red lines represent mirror symmetry planes parallel to the stacking direction. Finally, the green dot in the AB bilayer denotes the in-plane position of the two three-fold rotation axes parallel to the stacking direction, while the dashed green line in the AB$'$ bilayer corresponds to an in-plane C$_2$ rotation axis.}
    \label{fig:Fig_struct_BL_AB_AB1}
\end{figure}
In Figure \ref{fig:Fig_struct_BL_AB_AB1} we present the crystal structures of BL-C$_3$N with AB (upper panel) and AB$'$ (lower panel) stackings: yellow (light blue) spheres denote Carbon (Nitrogen) atoms and small (large) atoms are located on the upper (lower) layer, denoted from now on as $\mathrm{L}_1$ ($\mathrm{L}_2$).
For both stacking motifs, we have obtained an in-plane lattice parameter of 4.849 \AA, slightly smaller than that of the isolated monolayer (4.857 \AA); the interlayer distance 
(evaluated as the separation along $z$ between Carbon atoms with the same in-plane coordinates) has been found equal to 3.22 \AA\ for the AB and 3.21 \AA\  for the AB$'$ stacking. 
These values are in agreement with those obtained with PBE-D2 calculations in Ref.~[\onlinecite{Wei_2021}], while  
slightly smaller than the interlayer distances computed with VdW-functionals in Ref.~[\onlinecite{Wu_2018}]. 

We now briefly discuss the crystal symmetries of the two stackings. The point group of AB-C$_3$N is D$_{3\mathrm{d}}$, and also includes non-symmorphic operations. 
This stacking possesses a spatial inversion center (red dot in the upper panel of Fig.~\ref{fig:Fig_struct_BL_AB_AB1}) together with
a three-fold C$_3$ rotation axis along $z$-direction, 
whose in-plane position is denoted by a green dot in Fig.~\ref{fig:Fig_struct_BL_AB_AB1}. Furthermore, this stacking motif is invariant under mirror reflections $\sigma$  w.r.t. planes parallel to the stacking direction and represented by dashed red lines in Fig.~\ref{fig:Fig_struct_BL_AB_AB1}: These planes are respectively denoted as $\hat{\sigma}_{\Gamma \mathrm{M}}$, $\hat{\sigma}_{\Gamma \mathrm{M'}}$, and $\hat{\sigma}_{\Gamma \mathrm{M''}}$ as they are aligned to these high symmetry directions in the BZ.
AB$'$-C$_3$N has lower symmetry:
Its point group is C$_{2\mathrm{h}}$, 
that includes 
the inversion symmetry, a two-fold in-plane rotation axis lying between the two C$_3$N planes  (represented by the green dashed line in the lower panel of Fig.~\ref{fig:Fig_struct_BL_AB_AB1} and denoted as $\hat{\mathrm{C}}_{2}^{\Gamma \mathrm{K}}$), and a mirror symmetry plane $\hat{\sigma}_{\Gamma \mathrm{M''}}$ parallel to the $\Gamma \mathrm{M''}$ direction in the BZ.
\begin{figure*}
  \centering
    \includegraphics[width=1\textwidth]{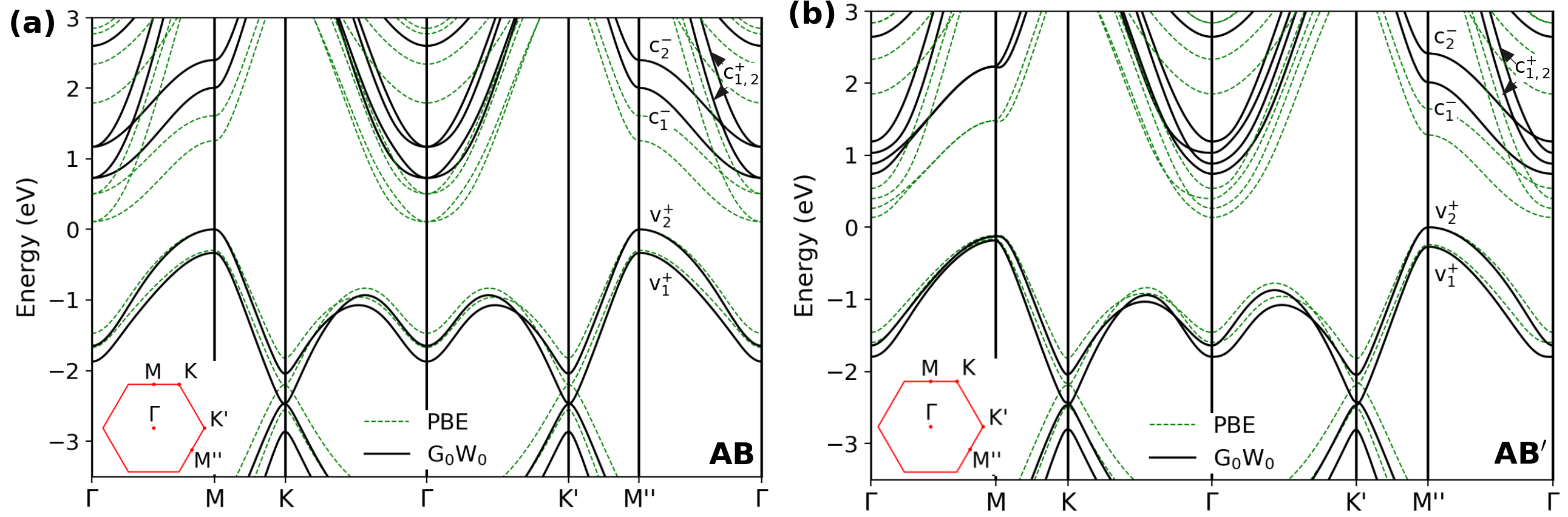}
    \caption{Electronic band structure computed at the DFT-PBE (green-dashed lines) and with G$_0$W$_0$ approximation (continuous black lines), for AB-C$_3$N (a) and AB$'$-C$_3$N (b). The insets represent the hexagonal Brillouin zone, together with the high symmetry points defining the paths where bands are computed. The parity of the topmost valence and the lowest conduction bands along $\Gamma \mathrm{M}''$ direction w.r.t. mirror symmetry $\hat{\sigma}_{\Gamma \mathrm{M''}}$ are indicated: the notation $n_{i}^{p}$ indicates that band $n_i$ has parity $p$, being $p=+(-)$ for even (odd) states and $n = \{v,c\}$. In both images, the top valence band energy is shifted to 0 eV.} 
    \label{fig:Fig_bands_BL_AB_AB1}
\end{figure*}

In Figure \ref{fig:Fig_bands_BL_AB_AB1} we present the electronic band structure of AB-C$_3$N (panel a) and AB$'$-C$_3$N (panel b), computed 
first 
within DFT-PBE (dashed green lines) and 
then 
including QP corrections at the G$_0$W$_0$ level (solid black lines). We note that the two stackings are characterized by an indirect band gap,
both at the DFT and GW level. In the case of AB stacking, the electronic band dispersion along $\Gamma$M and $\Gamma\mathrm{M''}$ coincide, as a consequence of the three-fold rotational symmetry. Therefore, the highest-energy valence band has six equivalent maxima, 
 while the bottom of the conduction band is located at the $\Gamma$ point. We also note that the presence of doubly degenerate bands at $\Gamma$ is consistent with irreducible representations of dimension 2 in the $\mathrm{D}_{3\mathrm{d}}$ point group. The calculated indirect gap at PBE level is 0.108 eV, which is increased to 0.72 eV once QP corrections are taken into account. Finally, the direct band gap is found along the $\Gamma\mathrm{M}$
direction, at 
\textbf{k} located approximately at half the $\Gamma\rightarrow\mathrm{M}$ path: the G$_0$W$_0$ gap is 1.85 eV, 0.73 eV larger than the value obtained at the PBE level (1.12 eV).

In the AB$'$ stacking, the maximum of the valence band is found at the $\mathrm{M}''$ point, while the lowest unoccupied conduction state is at $\Gamma$, as in the case of the AB stacking motif. Our calculations give an indirect gap of 0.136 eV at the DFT-PBE level, 
which is increased to  
0.73 eV 
with the inclusion of QP corrections. The obtained indirect gap for AB$'$-C$_3$N is slightly smaller than the one measured experimentally with STS on SiO$_2$-Si substrates in Ref.~[\onlinecite{Wei_2021}] (0.85$\pm$0.03 eV), but in better agreement than other GW calculations~\cite{Wu_2018} where the Hybertsen-Louie plasmon-pole model was used~\cite{Hybertsen_1986}. Because of the lack of three-fold rotational symmetry around the $z$ axis, the directions $\Gamma\mathrm{M}$ and $\Gamma\mathrm{M}''$ are no longer equivalent. As a result, the minimum direct gap is found approximately at half the $\Gamma\rightarrow\mathrm{M}''$ path, with a value around 1.79 eV (1.09 eV within DFT) while the indirect gap between the conduction at $\Gamma$ and the top-valence at $\mathrm{M}$ is slightly larger than the one between $\Gamma$ and $\mathrm{M}''$. We obtain a $\Gamma_c - \mathrm{M}_v$ gap of 0.87 eV (0.249 eV) at the G$_0$W$_0$ (DFT-PBE) level. Electronic gaps computed within the G$_0$W$_0$ approximation for both stacking motifs at high symmetry points are summarized in Table \ref{table:table_gaps}.

Comparing the band dispersions for AB and AB$'$ stackings, we notice that, for $\mathbf{k}$ along the $\Gamma\mathrm{M}$ direction, the lowest pair of conduction bands are almost degenerate (splitting of about 1 meV) in the AB$'$ stacking, while well separated in the AB case (splitting larger than 0.2 eV). 
In the \suppinfo, we provide a qualitative explanation of this peculiar lack of splitting among the two lowest conduction bands in AB$'$-C$_3$N, analysing the quasi-symmetries of the sublattice where conduction states are localized once $\mathbf{k}$ is taken along $\Gamma\mathrm{M}$ direction.
A similar behaviour is also observed for the two topmost valence states, where the splitting is, however, not negligible also in the AB$'$ stacking (splitting of about 50 meV). 
Overall, the obtained gaps for the AB and AB$'$ stacking are similar (in agreement with the hybrid-DFT results of Ref.~[\onlinecite{Wei_2021}]), while the electronic dispersions differ because of the different symmetry properties of the two stackings. 
We point out that 
there are clear differences with the monolayer case, as we observe a strong reduction in both the indirect and direct electronic gaps (see Table.~\ref{table:table_gaps} for a comparison). 

We complete our analysis
by recalling that both stacking configurations are invariant under mirror reflection $\hat{\sigma}_{\Gamma \mathrm{M''}}$, therefore electronic states for \textbf{k} along this direction can be classified in terms of their parity w.r.t. such symmetry. We have numerically found that, in both stackings, the two highest occupied valence bands are even w.r.t. $\hat{\sigma}_{\Gamma \mathrm{M''}}$, (here  denoted as $v_{i}^{+}$ in Fig.~\ref{fig:Fig_bands_BL_AB_AB1}, with $i = 1,2$). On the other hand, if \textbf{k} has modulus in the range $[\frac{|\Gamma\mathrm{M}|}{3},|\Gamma\mathrm{M}|]$ the two lowest unoccupied conduction bands are $\hat{\sigma}_{\Gamma \mathrm{M''}}$-odd ($c_{i}^{-}$). 
Instead, 
for \textbf{k} close to $\Gamma$, the second and the third  conduction bands are 
even w.r.t  $\hat{\sigma}_{\Gamma \mathrm{M''}}$ and  exhibit a stronger dispersion with \textbf{k}, if compared with odd conduction states. We point out that a similar analysis can be carried out also for electronic states along this direction in monolayer C$_3$N. In that case, the  
highest 
valence (lowest  conduction) is even (odd) w.r.t. the mirror symmetry along $\Gamma\mathrm{M}''$ direction, while the second  conduction is even.
In the case of AB stacking, the same parity analysis can be presented for the bands along the direction $\Gamma\mathrm{M}$, as the crystal is invariant w.r.t. $\hat{\sigma}_{\Gamma \mathrm{M}}$ mirror symmetry.
Such symmetry classification will be exploited to understand bilayer optical properties in the following.
\begin{table}
    \centering
    \begin{ruledtabular}
    \begin{tabular}{c|ccc}
      {\bf Gaps} & {\bf ML} Ref.~[\onlinecite{Miki_2022}]
      & {\bf AB} & {\bf AB$'$} \\[3pt]
      \hline & \\[-1ex]
      $\Gamma_c - \Gamma_v$             & 2.96 &  2.36 & 2.39 \\[3pt]
      $\mathrm{M}_c - \mathrm{M}_v$     & 2.67 &   2.02  & 2.40  \\[3pt]
      $\mathrm{M}''_c - \mathrm{M}''_v$ &      &   2.02  & 2.03  \\[3pt]
      $\Gamma_c - \mathrm{M}_v$         & 1.42 &   0.72  & 0.87  \\[3pt]
      $\Gamma_c - \mathrm{M}''_v$       &      &   0.72  & 0.73 \\[3pt]
      Min. direct gap                   & 2.62 &   1.85  & 1.79 \\[3pt]
    \end{tabular}
    \end{ruledtabular}
    \caption{Direct and indirect gaps (in eV) at high symmetry points for AB and AB$'$ BL-C$_3$N, obtained at the G$_0$W$_0$ level, compared with monolayer (ML) data (G$_0$W$_0$ on top of DFT-PBE) from Ref.~[\onlinecite{Miki_2022}]. The last row indicates the minimum direct electronic gap evaluated along $\Gamma\mathrm{M}''$ direction.}
    \label{table:table_gaps}
\end{table}
%
%
%
%
%
\section{Optical absorption in bilayer C$_3$N with AB and AB$'$ stackings}
\label{sec:optics}
%
\begin{figure*}
  \centering
    \includegraphics[width=1\textwidth]{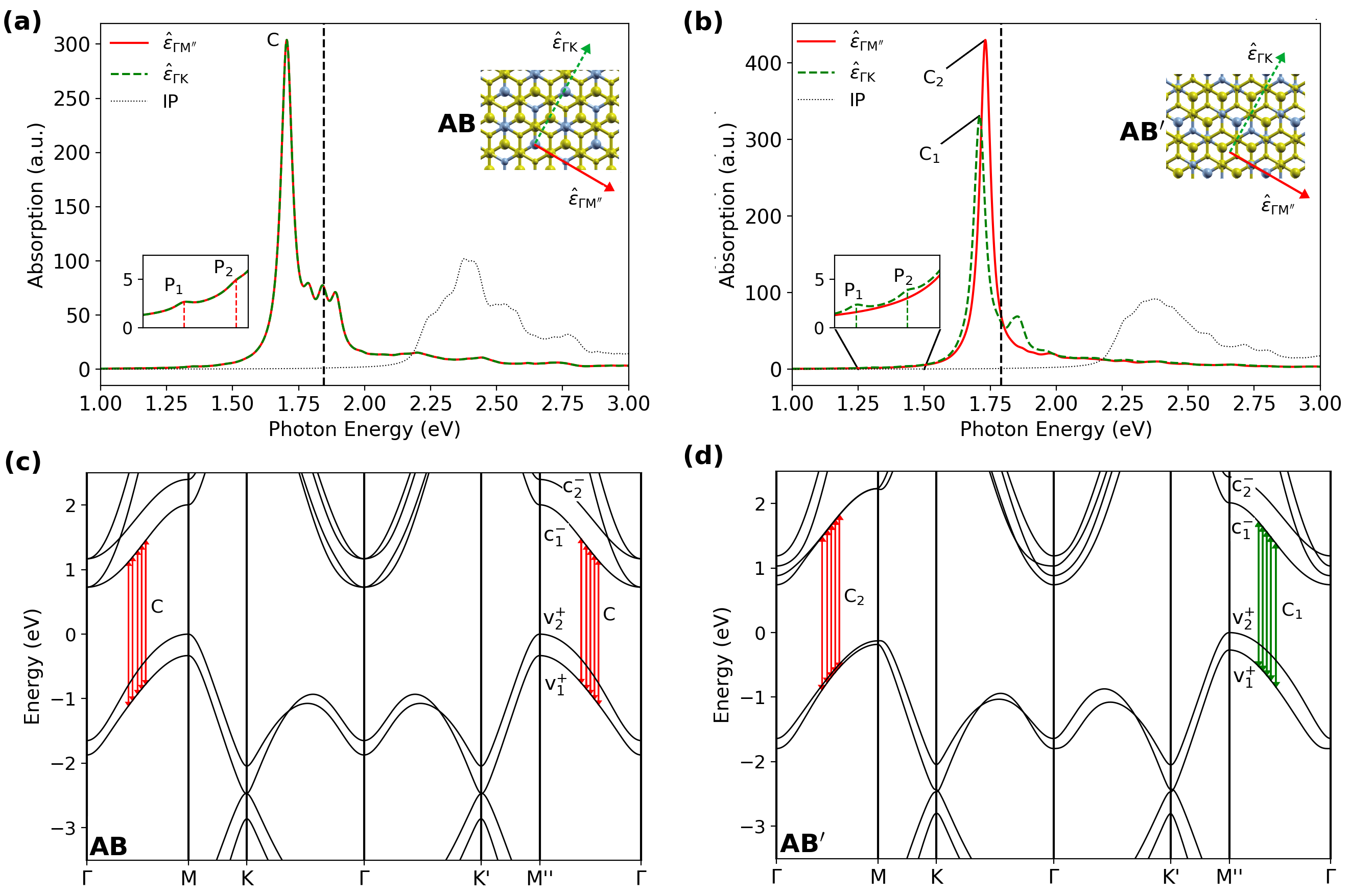}
    \caption{Optical absorption spectra for AB-C$_3$N (a) and AB$'$-C$_3$N (b). Solid red (dashed green) lines correspond to spectra computed with light polarization along the ${\Gamma\mathrm{M}}''$ (${\Gamma\mathrm{K}}$) direction, while dotted lines are the independent particle spectra evaluated for polarization along ${\Gamma\mathrm{M}}''$. The polarization directions are indicated on top of the crystal structures for clarity. The insets  highlight low energy 
    quasi-dark peaks labelled as P$_1$ and P$_2$, while the vertical black dashed lines indicate the position of the direct QP band gap obtained within G$_0$W$_0$. The      band structure at the G$_0$W$_0$ level 
     of AB-C$_3$N is shown in (c), with red arrows indicating the transitions mainly responsible for the C absorption peak in (a). Similarly, the single particle bands of AB$'$-C$_3$N are displayed in (d): green (red) arrows emphasize the transitions mostly involved in C$_\mathrm{1}$ (C$_\mathrm{2}$) peaks labelled in b.} 
    \label{fig:Fig_Total_abs}
\end{figure*}
We now turn to the discussion of the optical properties of bilayer C$_3$N.
In Figure~\ref{fig:Fig_Total_abs} we show the absorption spectra computed for AB (panel a) and AB$'$ bilayer C$_3$N (panel b), at the independent particle level (dotted black lines), and with the inclusion of electron-hole interaction by solving the BSE as detailed in Section~\ref{sec:methods}. Green dashed (solid red) lines have been obtained assuming light polarized along the ${\Gamma\mathrm{K}}$ (${\Gamma\mathrm{M}^{''}}$) direction. For clarity, polarization versors $\hat{\epsilon}$ are shown together with the crystal structures in the insets.

The AB spectrum is dominated by an intense peak (denoted as C) at energy $E\approx 1.70$ eV, whose spectral position and intensity are not dependent on the polarization direction. Such peak is due to a  set of almost degenerate eigenstates of H$_{exc}$, characterized by relevant contributions from single particle transitions between the valence band $v^+_1$ and the conduction state $c^-_1$ along $\Gamma\mathrm{M}$ equivalent directions. Among these, transitions with the highest weights are denoted by arrows on the band structure shown in Fig.~\ref{fig:Fig_Total_abs}(c). We 
point out that transitions between $v^+_2$ and $c^-_2$ for $\mathbf{k}$ along the same direction also contribute to this absorption peak, 
though 
with a smaller weight than $v^+_1 \rightarrow c^-_1$, and are omitted for clarity in Fig.~\ref{fig:Fig_Total_abs}(c). %
The situation is slightly different in the case of AB$'$ C$_3$N. Indeed, also for this stacking motif the absorption spectrum is dominated by a single intense peak, but its position in energy and its strength depend on the in-plane light polarization direction. In Fig.~\ref{fig:Fig_Total_abs}(b), we have labelled as C$_1$ the main peak at 1.71 eV obtained for light polarization along ${\Gamma\mathrm{K}}$ and as C$_2$ the absorption maximum at 1.73 eV found for ${\Gamma\mathrm{M}^{''}}$-polarized light. 
The C$_1$ peak is mainly due to $v^+_1 \rightarrow c^-_1$ for $\mathbf{k}$ along $\Gamma\mathrm{M}^{''}$ direction, as shown schematically by green arrows in Fig.~\ref{fig:Fig_Total_abs}(d), again 
with a smaller contribution coming from $v^+_2$ to $c^-_2$ transitions for the same $\mathbf{k}$ points (not shown in Fig.~\ref{fig:Fig_Total_abs}d).
On the other hand, the C$_2$ absorption peak comes from transitions between the two highest occupied valence states and the two lowest (quasi-degenerate) conduction bands along ${\Gamma\mathrm{M}}$ and ${\Gamma\mathrm{M}}'$ directions (see red arrows in Fig.~\ref{fig:Fig_Total_abs}d).

First, we note that the energy window found for strong optical absorption in both stackings is still in a good range for solar energy conversion, not much lower than that found for monolayer C$_3$N of 1.82 eV in Ref.~[\onlinecite{Miki_2022}]. Furthermore, the associated transitions are still in region favourable to electron-hole splitting, as in the monolayer. This can be particularly interesting, since one can expect the bilayer to have a more stable structure, compared to a monolayer, when deposited on a suitable electrode.

Second, and central to this work, the striking feature of bilayer C$_3$N optical spectra is 
the apparent absence of intense absorption peaks due to strongly bound excitonic states formed by single particle transitions close to the electronic direct gap. As shown in the insets of Fig.~\ref{fig:Fig_Total_abs}(a,b), within the energy range between 1.25 and 1.5 eV, 
two absorption structures (labelled in both cases as P$_1$ and P$_2$) are present, but they exhibit optical strengths which are almost two orders of magnitude smaller than the most intense peaks. 

%
\begin{figure}
  \centering
    \includegraphics[width=0.5\textwidth]{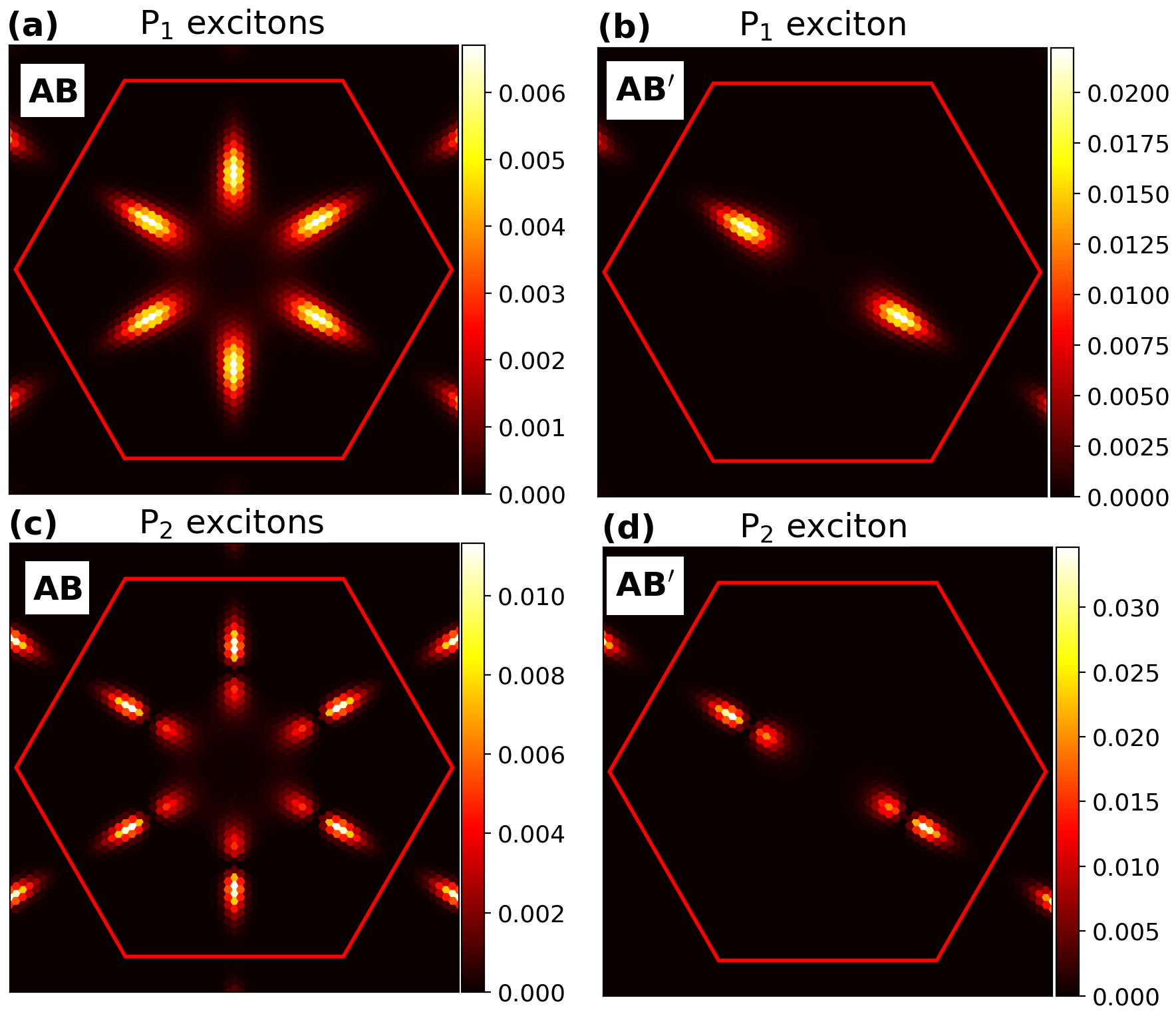}
    \caption{Contributions of single particle transitions between the highest occupied valence and the lowest unoccupied conduction in the BZ to excitons P$_1$ and P$_2$ in AB-C$_3$N (a and c) and in AB$'$-C$_3$N (b and d), as defined by Eq. \eqref{eq:eq_exc_k_space}.}
    \label{fig:Exc_A_and_B_reciprocal}
\end{figure}

In AB-C$_3$N, both P$_1$ and P$_2$ peaks are due to a pair of degenerate excitons, respectively at energies $E_{\mathrm{P}_{1}}$ = 1.35 eV and $E_{\mathrm{P}_{2}}$ = 1.47 eV, with oscillator strengths not dependent on the polarization direction. For completeness, we point out that diagonalization of the excitonic Hamiltonian also provides other excitonic resonances (the lowest with energy of 1.34 eV) which are characterized by null oscillator strength within numerical accuracy. Such excitons, dark by strict symmetry reasons, will not be considered further in the following. 
P$_1$ and P$_2$ excitons are almost totally composed by electron-hole transitions between the highest occupied valence $v^+_2$ and the lowest unoccupied conduction $c^-_1$ states, with wave-vectors $\mathbf{k}$ along $\Gamma\mathrm{M}$ and equivalent directions in the BZ. This is better clarified in Fig.~\ref{fig:Exc_A_and_B_reciprocal}(a,c), where we show plots of the quantity 
\begin{equation}
    A^{vc}(\mathbf{k}) = \sum_{\lambda}\big |A_\lambda(vc\mathbf{k})\big |^2
    \label{eq:eq_exc_k_space}
\end{equation}
for the P$_1$ and P$_2$ excitons, respectively. In Eq.~\eqref{eq:eq_exc_k_space}, the index $v$ ($c$) is fixed to the last valence (first conduction) band and the sum over $\lambda$ is performed over the pair of degenerate states responsible for the P$_1$ and P$_2$ peaks.
We notice that single particle transitions forming the excitons P$_1$ are mainly localized in the central region of $\Gamma\mathrm{M}$ and equivalent directions, with $A^{vc}(\mathbf{k})$ having non-negligible values for $|\mathbf{k}|$ mainly in the interval $[\frac{1}{3},\frac{2}{3}]|\Gamma\mathrm{M}|$. On the other hand, excitons P$_2$ (Fig. \ref{fig:Exc_A_and_B_reciprocal}c) are still localized along $\Gamma\mathrm{M}$ directions, but the corresponding function $A^{vc}(\mathbf{k})$ has intense contributions from transitions slightly closer to the $\mathrm{M}$ point and exhibits a node for $\mathbf{k}$ points along this direction.

In the case of the AB$'$ stacking, P$_1$ and P$_2$ peaks are related each to a single-nondegenerate exciton at energies $E_{\mathrm{P}_1}$ = 1.30 eV and $E_{\mathrm{P}_2}$ = 1.42 eV, respectively. Differently from the AB case, the oscillator strength of these excitations is polarization dependent as shown in 
the inset of Fig.~\ref{fig:Fig_Total_abs}(b), where we notice that both resonances are dark for light polarization along the ${\Gamma\mathrm{M}}''$ direction while they exhibit a small but non-zero optical activity for incoming light polarized along ${\Gamma\mathrm{K}}$. As in the AB stacking, such excitons are mainly formed by transitions between the highest occupied valence and the lowest conduction band. In Fig.~\ref{fig:Exc_A_and_B_reciprocal}(b,d) we show the functions $A^{vc}(\mathbf{k})$ computed for exciton P$_1$ and P$_2$, respectively. We see that, in both cases, the transitions involved in these excitations are strongly localized along the single $\Gamma\mathrm{M}''$ direction of the BZ (see e.g. Fig.~\ref{fig:Fig_bands_BL_AB_AB1}), with the P$_1$ exciton having the main contributions coming from the middle of $\Gamma\mathrm{M}''$, where the minimum direct electronic band gap is located, and the P$_2$ resonance characterized by a node along this BZ-line.

The observed polarization dependence in the optical absorption of AB$'$-C$_3$N can be rationalized via symmetry arguments, 
similarly to the analysis proposed in Ref.~[\onlinecite{Paleari_2018}] for bilayer hBN. 
The point group of AB$'$-C$_3$N is C$_{\mathrm{2h}}$, so the in-plane exciton dipole operator projected along ${\Gamma\mathrm{M}}''$ (${\mathbf{D}}_{\Gamma\mathrm{M}''}$)  transforms as the B$_{u}$ irreducible representation of C$_{\mathrm{2h}}$, while ${\mathbf{D}}_{\Gamma\mathrm{K}}$  as A$_u$. 
In fact, ${\mathbf{D}}_{\Gamma\mathrm{M}''}$ is invariant under the mirror symmetry $\hat{\sigma}_{\Gamma\mathrm{M}''}$ (as oriented along $\Gamma\mathrm{M}''$), but it changes sign under inversion and $\hat{C}^{\Gamma\mathrm{K}}_2$ rotation (see Fig.~\ref{fig:Fig_struct_BL_AB_AB1}). Differently, ${\mathbf{D}}_{\Gamma\mathrm{K}}$ is even w.r.t. $\hat{C}^{\Gamma\mathrm{K}}_2$ and odd under $\hat{\sigma}_{\Gamma\mathrm{M}''}$ and inversion, therefore behaving as the $A_u$(C$_{2h}$) irreducible representation.
Therefore, the eigenstates $|\lambda\rangle$ of $H_{exc}$ such that $\langle 0|\mathbf{D}|\lambda\rangle \neq 0$ ($|0\rangle$ being the excitonic vacuum transforming as the fully symmetric representation A$_g$)  transform as A$_u$ or B$_u$ if ${\mathbf{D}}$ is projected along  ${\Gamma\mathrm{K}}$ or ${\Gamma\mathrm{M}}''$, respectively. Considering for example A$_u$ states, these will have null optical activity by symmetry, once incoming light is polarized along ${\Gamma\mathrm{M}}''$. Therefore, we can explain the presence of absorption peaks in AB$'$-C$_3$N which turn on and off according to the chosen polarization direction. Furthermore, following the presented analysis, we can assign P$_1$ and P$_2$ excitons to the A$_u$ representation, exactly as the eigenstates of H$_{exc}$ responsible for the C$_1$ peak, while C$_2$ is expected to be due to excitations transforming as B$_u$.

The situation is different for AB stacked C$_3$N. For simplicity, we discuss the brightness of excitonic eigenstates using the subgroup C$_{3v}$ of D$_{3d}$, formed by the mirror planes $\sigma$ along $\Gamma\mathrm{M}$ directions together with the two 3-fold rotations depicted in the upper panel of Fig.~\ref{fig:Fig_struct_BL_AB_AB1}. The in-plane exciton dipole operator $\mathbf{D}$ transforms as the two-dimensional irreducible representation E of C$_{3v}$. As a consequence, all bright excitons behave as E of C$_{3v}$, so that they are expected to be  double-degenerate and characterized by an isotropic oscillator strength, in agreement with the numerical results obtained from \textit{ab initio} calculations.

\begin{figure}
  \centering
    \includegraphics[width=0.5\textwidth]{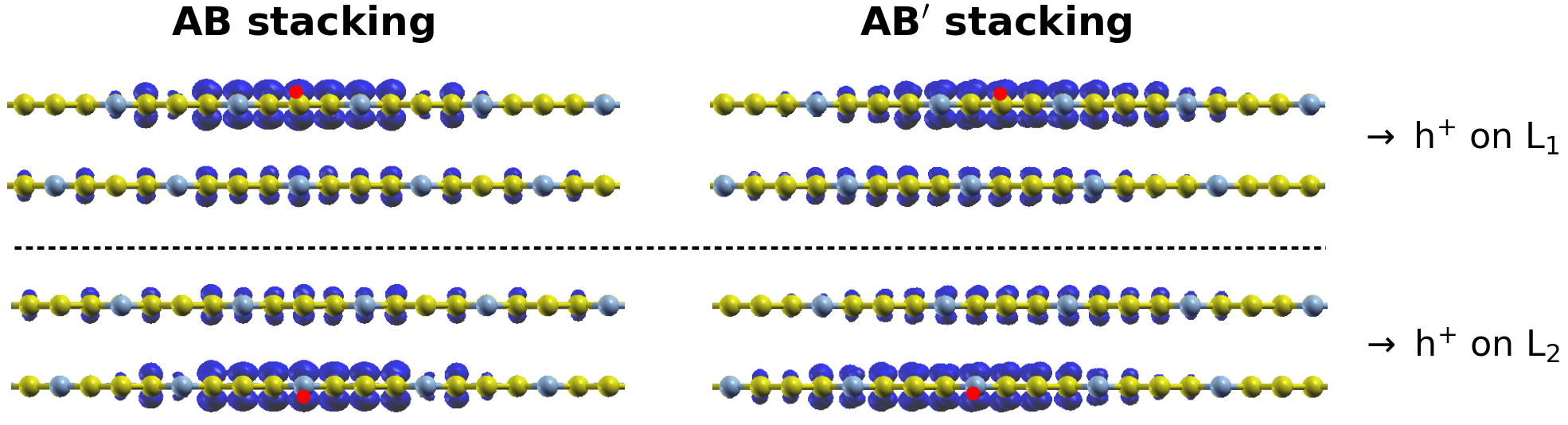}
    \caption{Wavefunctions for exciton P$_1$ computed in real space for AB (left) and AB$'$ (right) stackings. In each panel, the upper (lower) wavefunction has been computed assuming the hole localized on a Nitrogen atom on layer L$_1$ (L$_2$). The hole position along the stacking direction is indicated by the red circle.}
    \label{fig:Real_space_A_exc}
\end{figure}

We now turn our attention to the quasi-dark nature of the low-lying bound excitons P$_1$ and P$_2$ in both the considered stackings. We point out that such small optical activity cannot be related to an interlayer nature of these excitons, i.e. it is not due to a negligible spatial overlap between electron and hole wavefunctions. This point is better clarified by looking, for example, at the real space wavefunctions of exciton P$_1$, as shown in Fig.~\ref{fig:Real_space_A_exc}, with AB (AB$'$) results reported in the left (right) panel. 
For each stacking, the upper (lower) wavefunction has been obtained assuming the hole (represented by the red dot) fixed on layer L$_1$ (L$_2$) and located in the plane close to a nitrogen atom.
The wavefunctions clearly indicate the intralayer nature of such excitons: In fact, looking at the excitonic wavefunction isosurfaces, we notice that the electron has a high probability to be found on the same layer on which the hole is localized. 

Therefore, the negligible dipole strength of these low-lying excitons is a consequence of the small interband dipole associated to the electron-hole single particle transitions involved in these excitations. This can be understood from the independent particle (IP) absorption spectrum shown in Fig.~\ref{fig:Fig_Total_abs}(a,b) as dotted lines, where, 
in both cases, the optical signal is negligible for energies close to the direct QP gap, with the IP absorption onset located at higher photon energies. In the following section we discuss in greater detail the single particle states involved in these low-lying excitons and we propose a possible rationale for the observed IP absorption quenching.
%
%
%
\section{Rationale for quenching of low energy absorption in AB and AB$'$ BL-C$_3$N}
\label{sec:INTERBAND_DIPOLE}
%
%
In this Section we develop a model for the electronic bands in proximity of the direct gap to rationalize the small oscillator strength of the single particle transitions close to the direct electronic gap. As the following analysis is valid for both stackings, here we focus on the case of AB$'$-C$_3$N (results for AB motif are presented in the \suppinfo). We will restrict our analysis to $\mathbf{k}$ along $\Gamma\mathrm{M}''$, in the region where valence to conduction transitions giving the highest contribution to the low-energy quasi-dark excitons are located.

As already discussed in the literature~\cite{Wei_2021,Wu_2018}, the lowest lying conduction bands and the highest valence states have a $\pi$ character. therefore, we can analyse them using a tight-binding (TB) Hamiltonian, obtained considering one 2p$_z$ orbital for each atom. In the following we will denote as $\mathbf{\tau}_{\alpha}$ the position of both the atom $\alpha$ and the 2p$_z$ orbital localized on it. In practice, we construct a TB Hamiltonian as
\begin{equation}
    H^{\mathrm{2L}}_{\alpha,\beta}(\mathbf{k}) = \sum_{\mathbf{R}}e^{i\mathbf{k}\cdot\mathbf{R}} \,t(\alpha 0,\beta\mathbf{R}),
    \label{eq_TB_H}
\end{equation}
where $\mathbf{R}$ is a lattice vector and $t(\alpha 0,\beta\mathbf{R}) = \langle\alpha0|{H}|\beta\mathbf{R}\rangle$ are the hopping matrix elements between two 2p$_z$ orbitals, localized at sites $\mathbf{\tau}_{\alpha}$ and $\mathbf{\tau}_{\beta} + \mathbf{R}$, respectively. These matrix elements have been computed fully \textit{ab initio} by Wannierization~\cite{Souza_2001,Marzari_2012} of DFT bands using the {\tt{Wannier90}}~\cite{MOSTOFI_2008,MOSTOFI_2014,Pizzi_2020} code. Details about the procedure are provided in the \suppinfo.

As there are 16 atoms in the unit cell, at a general $\mathbf{k}$ the TB Hamiltonian can be written as a 16$\times$16 Hermitian matrix, in a block-like form as
\begin{equation}
    \mathbf{H}^{\mathrm{2L}}(\mathbf{k}) =
    \begin{bmatrix}
          \mathbf{H}^{\mathrm{L}_1}(\mathbf{k}) & \mathbf{H}^{\mathrm{IL}}(\mathbf{k})\\
          \mathbf{H}^{\mathrm{IL}}(\mathbf{k})^\dagger & \mathbf{H}^{\mathrm{L}_2}(\mathbf{k})
    \end{bmatrix},
    \label{eq:BLOCK_TB_H}
\end{equation}
where $\mathbf{H}^{\mathrm{L}_1}$, $\mathbf{H}^{\mathrm{L}_2}$, and $\mathbf{H}^{\mathrm{IL}}$ are 8$\times$8 
matrices corresponding to the different layers and their coupling. Indeed, to obtain this separation, we have grouped together the orbitals localized on L$_1$ and on L$_2$, associating to each orbital $\alpha_1$ localized at $\mathbf{\tau}_{\alpha_1}$ on layer L$_1$ an orbital $\alpha_2$ at $\mathbf{\tau}_{\alpha_2}$ on layer L$_2$, such that $\mathbf{\tau}_{\alpha_2} = \hat{\mathrm{I}}\mathbf{\tau}_{\alpha_2}$, $\hat{\mathrm{I}}$ being the inversion symmetry operator. Then $\mathbf{H}^{\mathrm{L}_1}$ ($\mathbf{H}^{\mathrm{L}_2}$) is the block of $\mathbf{H}^{\mathrm{2L}}$ which contains the intralayer hopping within layer L$_1$ (L$_2$), while $\mathbf{H}^{\mathrm{IL}}$ depends on the hopping integrals between orbitals on different layers. We can now write $\mathbf{H}^{\mathrm{2L}}(\mathbf{k}) = \mathbf{h}_{\mathrm{IN}}(\mathbf{k}) + \mathbf{h}_{\mathrm{IL}}(\mathbf{k})$, where $\mathbf{h}_{\mathrm{IN}}(\mathbf{k})$  is block-diagonal while $\mathbf{h}_{\mathrm{IL}}(\mathbf{k})$  is purely off-diagonal. We note that the subscripts "IN" and "IL"  stand for intralayer and interlayer, respectively.
Using spatial ($\hat{\mathrm{I}}$) and time ($\hat{\mathrm{T}}$) inversion symmetries, in Appendix \ref{sec:AppendixB} we show that, for each $\mathbf{k}$, $\mathbf{h}_{\mathrm{IN}}(\mathbf{k})$ has a spectrum of eigenvalues $\varepsilon_{n\mathbf{k}}^0$ which are degenerate in pairs. Furthermore, we also show that the eigenspace associated to each eigenvalue $\varepsilon_{n\mathbf{k}}^0$ is spanned by two Bloch states $|\phi^{\mathrm{L}_1}_{n\mathbf{k}}\rangle$ and $|\phi^{\mathrm{L}_2}_{n\mathbf{k}}\rangle$ such that $|\phi^{\mathrm{L}_2}_{n\mathbf{k}}\rangle =- \hat{\mathrm{I}}\cdot\hat{\mathrm{T}}|\phi^{\mathrm{L}_1}_{n\mathbf{k}}\rangle$ and with $|\phi^{\mathrm{L}_i}_{n\mathbf{k}}\rangle$ localized on layer $\mathrm{L}_i$.

Before proceeding, we clarify the physical meaning of the splitting of $\mathbf{H}^{\mathrm{2L}}$ into $\mathbf{h}_{\mathrm{IN}}$ and $\mathbf{h}_{\mathrm{IL}}$. In particular, $\mathbf{h}_{\mathrm{IN}}$ can be thought as an intralayer Hamiltonian, 
describing the two layers as not interacting with each other. Note however that the presence of the other layer is implicitly considered as the intralayer matrix elements in $\mathbf{H}^{\mathrm{L}_1}$ ($\mathbf{H}^{\mathrm{L}_2}$) are affected by the presence of opposite layer L$_2$ (L$_1$). On the other hand, $\mathbf{h}_{\mathrm{IL}}$ describes the interlayer interaction and acts as a perturbation of $\mathbf{h}_{\mathrm{IN}}$ since interlayer hopping integrals are typically smaller than the intralayer ones.
With this interpretation, the states $|\phi^{\mathrm{L}_1}_{n\mathbf{k}}\rangle$ and $|\phi^{\mathrm{L}_2}_{n\mathbf{k}}\rangle$ can be thought as Bloch states with the same energy $\varepsilon_{n\mathbf{k}}^0$ localized on one of the two monolayers, 
 if the coupling $\mathbf{h}_{IL}$ is set to zero. We now define $\varepsilon_{v\mathbf{k}}^0$ ($\varepsilon_{c\mathbf{k}}^0$) as the energy of the highest occupied valence (lowest unoccupied conduction) band on these two non-interacting layers. The effect of the interlayer coupling will be to mix these layer-localized wavefunctions, in order to give the electronic states of the bilayer. 

While the discussion presented so far is general and valid for each $\mathbf{k}$-point of the BZ, we now specialize to $\mathbf{k}$-vectors along the $\Gamma\mathrm{M}''$ direction. 
For these wavevectors, $\mathbf{H}^{\mathrm{2L}}$ commutes with $\hat{\sigma}_{\Gamma\mathrm{M}''}$ and the same is 
valid for $\mathbf{h}_{\mathrm{IN}}$ and $\mathbf{h}_{\mathrm{IL}}$ separately. Therefore,
\begin{equation}
    \begin{split}
        \hat{\sigma}_{\Gamma\mathrm{M}''}|\phi^{\mathrm{L}_i}_{v\mathbf{k}}\rangle &= |\phi^{\mathrm{L}_i}_{v\mathbf{k}}\rangle, \\
        \hat{\sigma}_{\Gamma\mathrm{M}''}|\phi^{\mathrm{L}_i}_{c\mathbf{k}}\rangle &= -|\phi^{\mathrm{L}_i}_{c\mathbf{k}}\rangle,
    \end{split}
    \label{eq:eq_sigma_model}
\end{equation}
with $i = 1,2$ (we have also numerically verified these relations by computing the eigenstates of $\mathbf{h}_{\mathrm{IN}}$).
We now include the effect of $\mathbf{h}_{\mathrm{IL}}$ using first order degenerate perturbation theory, separately diagonalizing the matrix representation of ${h}_{\mathrm{IL}}$ on the two subspaces $\{|\phi^{\mathrm{L}_1}_{v\mathbf{k}}\rangle,|\phi^{\mathrm{L}_2}_{v\mathbf{k}}\rangle   
\}$ and $\{|\phi^{\mathrm{L}_1}_{c\mathbf{k}}\rangle,|\phi^{\mathrm{L}_2}_{c\mathbf{k}}\rangle   
\}$. With this procedure, we obtain the two highest-energy ($\hat{\sigma}_{\Gamma\mathrm{M}''}$-even) valence bands  $\{|\varphi_{v_1\mathbf{k}}\rangle,|\varphi_{v_2\mathbf{k}}\rangle 
\}$ and the two lowest-energy ($\hat{\sigma}_{\Gamma\mathrm{M}''}$-odd) conduction states $\{|\varphi_{c_1\mathbf{k}}\rangle,|\varphi_{c_2\mathbf{k}}\rangle \}$ in the bilayer. Such states have been labelled as ($v^+_1$,$v^+_2$) and ($c^-_1$,$c^-_2$), respectively, in Fig.~\ref{fig:Fig_bands_BL_AB_AB1}(b).
They can be compactly written as 
\begin{eqnarray}
    |\varphi_{n_j\mathbf{k}}\rangle &=& \frac{1}{\sqrt{2}}
    \bigg [|\Tilde{\phi}^{\mathrm{L}_1}_{n\mathbf{k}}\rangle + s_j|\Tilde{\phi}^{\mathrm{L}_2}_{n\mathbf{k}}\rangle \bigg ]
    \label{eq:pert_theory_GM}
    \\
    E_{n_j\mathbf{k}} &=& \varepsilon_{n,\mathbf{k}}^0 + s_j|\Delta_{n\mathbf{k}}| .
    \label{eq:eq_ENE_pert}
\end{eqnarray}
In Eqs.~\eqref{eq:pert_theory_GM}-\eqref{eq:eq_ENE_pert}, $n = v,c$, $j = 1,2$, $s_j = -1$ ($+1$) for $j=1$ ($j=2$), $\Delta_{n\mathbf{k}} = \langle\phi^{\mathrm{L}_1}_{n\mathbf{k}}|{h}_{\mathrm{IL}}|\phi^{\mathrm{L}_2}_{n\mathbf{k}}\rangle$, and
\begin{equation}
    \begin{split}
        |\Tilde{\phi}^{\mathrm{L}_1}_{n\mathbf{k}}\rangle &= e^{+i\frac{\gamma_{n\mathbf{k}}}{2}}|\phi^{\mathrm{L}_1}_{n\mathbf{k}}\rangle,\\[5pt]
        |\Tilde{\phi}^{\mathrm{L}_2}_{n\mathbf{k}}\rangle &= e^{-i\frac{\gamma_{n\mathbf{k}}}{2}}|\phi^{\mathrm{L}_2}_{n\mathbf{k}}\rangle,
    \end{split}
    \label{eq:tilde_states_eq}
\end{equation}
where $\gamma_{n\mathbf{k}} = \mathrm{Arg}[\Delta_{n\mathbf{k}}]$ guarantees that the relative phase between the projections $c_\alpha(n_{j}\mathbf{k})=\langle \alpha \mathbf{k} |\varphi_{n_{j}\mathbf{k}}\rangle$  of a Bloch state $|\varphi_{n_{j}\mathbf{k}}\rangle$ on 2p$_z$ orbitals localized on different layers is gauge invariant, i.e. it does not change under the transformation $c_{\alpha_1}(n\mathbf{k}) \rightarrow e^{i\eta}c_{\alpha_1}(n\mathbf{k})$, $\eta$ being an arbitrary phase.
See App.~\ref{sec:AppendixB} for further details.

We point out that the zero-order expression given by Eq.~\eqref{eq:pert_theory_GM} is a good approximation for the two lowest $\hat{\sigma}_{\Gamma\mathrm{M}''}$-odd conduction bands and the two highest-energy $\hat{\sigma}_{\Gamma\mathrm{M}''}$-even valence bands, for the considered $\mathbf{k}$ vectors along $\Gamma\mathrm{M}''$ direction. In principle one should also consider other terms in the expression of the eigenstates\cite{sakurai_napolitano_2017}, coming from higher orders of the perturbative series, describing the coupling between $|\phi^{\mathrm{L}_i}_{n\mathbf{k}}\rangle$ and the eigenstates of $\mathbf{h}_{\mathrm{IN}}$ with different eigenvalues. For $\mathbf{k}$ points around the middle of the $\Gamma\mathrm{M}''$ direction, these terms can be  neglected in first approximation, as the other eigenstates of $\mathbf{h}_{\mathrm{IN}}$ with the same $\hat{\sigma}_{\Gamma\mathrm{M}''}$-parity of $|\phi^{\mathrm{L}_i}_{n\mathbf{k}}\rangle$ have energies far from $\varepsilon_{n\mathbf{k}}^0$ (w.r.t. to the interlayer coupling strength), so that the hybridization is negligible. Our numerical results (not shown) indicate that these neglected terms become more relevant for $\mathbf{k}$ along the same direction, but closer to the $\Gamma$ point.
Therefore, looking at Eq.~\eqref{eq:pert_theory_GM}, we can understand that the lowest odd-conduction band and the highest even-valence can be seen, respectively, as antibonding and bonding combinations of the conduction and the valence states localized on the two monolayers. We recall that the states defined in Eq.~\eqref{eq:tilde_states_eq} are still eigenstates of the intralayer Hamiltonian $\hat{h}_{\mathrm{IN}}$, with $|\Tilde{\phi}^{\mathrm{L}_2}_{n\mathbf{k}}\rangle =- \hat{\mathrm{I}}\cdot\hat{\mathrm{T}}|\Tilde{\phi}^{\mathrm{L}_1}_{n\mathbf{k}}\rangle$.

\begin{figure*}
  \centering
    \includegraphics[width=0.95\textwidth]{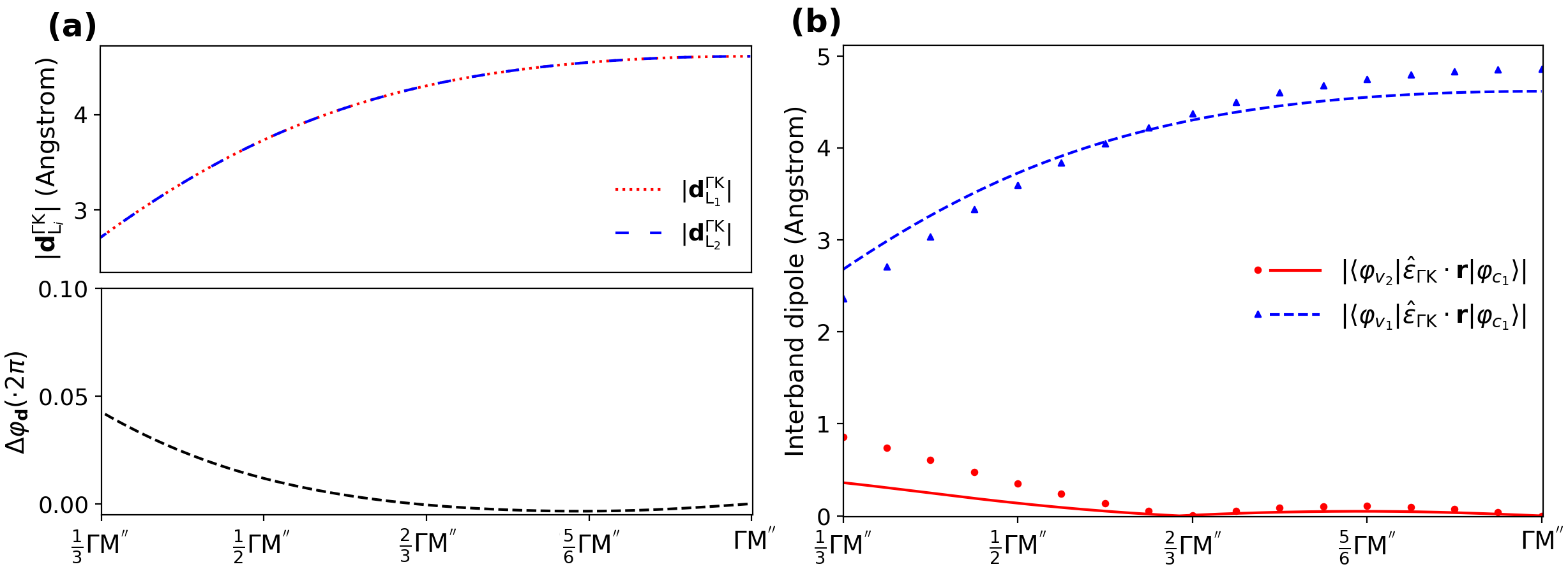}
    \caption{Interband dipole matrix elements for $\mathbf{k}$ along $\Gamma\mathrm{M}''$ direction, in the interval $[\frac{|\Gamma\mathrm{M}''|}{3},|\Gamma\mathrm{M}''|]$. In the upper panel of (a) we show the moduli of $d^{\epsilon}_{L_i}$ for light polarization along $\Gamma\mathrm{K}$ for the two layers $i = 1,2$, while in the lower panel the phase difference among $d^{\epsilon}_{L_1}$ and $d^{\epsilon}_{L_2}$ is presented. In (b) the red continuous (blue dashed) line corresponds to the interband dipole between $|\varphi_{v_2\mathbf{k}}\rangle$ ($|\varphi_{v_1\mathbf{k}}\rangle$) and $|\varphi_{c_1\mathbf{k}}\rangle$  computed using the TB model. Dots and triangles represent the same quantities computed fully \textit{ab initio} using {\tt{Yambo}} code. Light polarization versor is assumed aligned along $\Gamma\mathrm{K}$ direction.}
    \label{fig:Fig_dipoles_AB1}
\end{figure*}

Starting from Eq.~\eqref{eq:pert_theory_GM}, we can evaluate the interband matrix element between the last occupied valence $|\varphi_{v_2\mathbf{k}}\rangle$ and the lowest unoccupied conduction $|\varphi_{c_1\mathbf{k}}\rangle$ as
\begin{eqnarray}
    \nonumber
        d^{\epsilon}_{v_2c_1\mathbf{k}} = \frac{1}{2}\!\!\!\!\!&&\bigg[\langle\Tilde{\phi}^{\mathrm{L}_1}_{v\mathbf{k}}|\hat{\epsilon}\cdot\mathbf{r}|\Tilde{\phi}^{\mathrm{L}_1}_{c\mathbf{k}}\rangle - \langle\Tilde{\phi}^{\mathrm{L}_2}_{v\mathbf{k}}|\hat{\epsilon}\cdot\mathbf{r}|\Tilde{\phi}^{\mathrm{L}_2}_{c\mathbf{k}}\rangle\\
        &&+\langle\Tilde{\phi}^{\mathrm{L}_2}_{v\mathbf{k}}|\hat{\epsilon}\cdot\mathbf{r}|\Tilde{\phi}^{\mathrm{L}_1}_{c\mathbf{k}}\rangle - \langle\Tilde{\phi}^{\mathrm{L}_1}_{v\mathbf{k}}|\hat{\epsilon}\cdot\mathbf{r}|\Tilde{\phi}^{\mathrm{L}_2}_{c\mathbf{k}}\rangle\bigg ],
    \label{eq:eq_dipole}
\end{eqnarray}
where $\hat{\epsilon}$ is the light polarization. To make the treatment simpler, in the following we neglect the last two terms in Eq.~\eqref{eq:eq_dipole}, as they involve states localized on different layers, so that their value is generally small as a result of the reduced overlap among the wavefunctions. 
In this way, we obtain that the interband dipole is the difference between intralayer-interband dipoles:
\begin{equation}
    d_{\mathrm{L}_i}^{\epsilon}(\mathbf{k}) =\langle\Tilde{\phi}^{\mathrm{L}_i}_{v,\mathbf{k}}|\hat{\epsilon}\cdot\mathbf{r}|\Tilde{\phi}^{\mathrm{L}_i}_{c,\mathbf{k}}\rangle .
    \label{eq:intra_inter_dipole}
\end{equation}
The approach adopted to compute these matrix elements starting from the proposed tight binding model is discussed in the \suppinfo.

If $\hat{\epsilon}$ is chosen along the $\Gamma\mathrm{M}''$ direction, we immediately find that $d_{\mathrm{L}_1}^{\epsilon}(\mathbf{k}) = d_{\mathrm{L}_2}^{\epsilon}(\mathbf{k}) = 0$, because of the parity of layer-resolved states, given by Eq.~\eqref{eq:eq_sigma_model}. However, if $\hat{\epsilon}$ is taken along the direction $\Gamma\mathrm{K}$, orthogonal to $\Gamma\mathrm{M}''$, we cannot explain the quenching by straightforward symmetry arguments. To clarify this point, in Fig.~\ref{fig:Fig_dipoles_AB1}(a) we show the modulus (upper panel) of the intralayer-interband dipoles $d_{\mathrm{L}_1}^{\Gamma\mathrm{K}}(\mathbf{k})$ and $d_{\mathrm{L}_2}^{\Gamma\mathrm{K}}(\mathbf{k})$, together with their relative phase (lower panel) $\Delta\varphi_{d}(\mathbf{k}) = \mathrm{Arg}[d_{\mathrm{L}_1}^{\Gamma\mathrm{K}}(\mathbf{k})] - \mathrm{Arg}[d_{\mathrm{L}_2}^{\Gamma\mathrm{K}}(\mathbf{k})]$.
Our results indicate that the obtained intralayer dipole matrix elements are equal in modulus 
and they exhibit a relative phase close to zero in the range of $\mathbf{k}$ points here considered, i.e. $|\mathbf{k}|$ in $[\frac{|\Gamma\mathrm{M}''|}{3},|\Gamma\mathrm{M}''|]$. As the total interband dipole is the difference among intralayer contributions -- see Eq.~\eqref{eq:eq_dipole} --, it will be almost zero, as a consequence of the destructive interference of the two layer-resolved components. In other words, the transition probability from $|\varphi_{v_2\mathbf{k}}\rangle$ to $|\varphi_{c_1\mathbf{k}}\rangle$ due to $\Gamma\mathrm{K}$-polarized light can be interpreted as the quantum superposition of the interband scattering processes occurring on the two layers separately, whose probability amplitudes are out-of-phase, giving an overall negligible interband oscillator strength.

In Fig.~\ref{fig:Fig_dipoles_AB1}(b), the continuous red line indicates the interband dipole between $|\varphi_{v_2\mathbf{k}}\rangle$ and $|\varphi_{c_1\mathbf{k}}\rangle$, computed using the perturbative solution of the TB model (see \suppinfo{} for details), while the red dots are the same quantities obtained \textit{ab initio} using {\tt{Yambo}}, to check the validity of our approximate treatment. We notice that, as this cancellation is not symmetry-constrained, the interband dipole is small, but not exactly zero. Such cancellation is exact at $\mathrm{M}''$ point, because of symmetry reasons. In fact, as $\mathrm{M}''$ is invariant under spatial inversion $\hat{\mathrm{I}}$, we can assign inversion-parity labels to the states at this point. \textit{Ab initio} results indicate that both $|\varphi_{v_2\mathbf{k}}\rangle$ and $|\varphi_{c_1\mathbf{k}}\rangle$ are odd under $\hat{\mathrm{I}}$ exactly as $\mathbf{d}_{\Gamma\mathrm{K}}$, therefore the overall matrix element is zero.  

We point out that the interband dipole is non-zero once the transition between a pair of bonding or antibonding combinations is considered. For example, taking into account the scattering  $|\varphi_{v_1\mathbf{k}}\rangle\rightarrow|\varphi_{c_1\mathbf{k}}\rangle$, the intralayer-interband transition amplitudes sum constructively giving an intense overall interband dipole. The intense optical activity of these transitions is responsible for the main absorption peak denoted as C$_1$ in Fig.~\ref{fig:Fig_Total_abs}. Such prediction is confirmed by data in Fig.~\ref{fig:Fig_dipoles_AB1}(b), where this quantity is shown as computed using the model (dashed blue line) and fully \textit{ab initio} using {\tt{Yambo}} (blue triangles). We notice that the reasonable agreement between the model and the \textit{ab initio} results is an \textit{a posteriori} confirmation of the validity of Eq.~\eqref{eq:pert_theory_GM} to describe single particle states along $\Gamma\mathrm{M}''$.

%
%
%
\section{The case of AA$'$ stacking}
\label{sec:STACKING_AA1}
\begin{figure}
  \centering
    \includegraphics[width=0.45\textwidth]{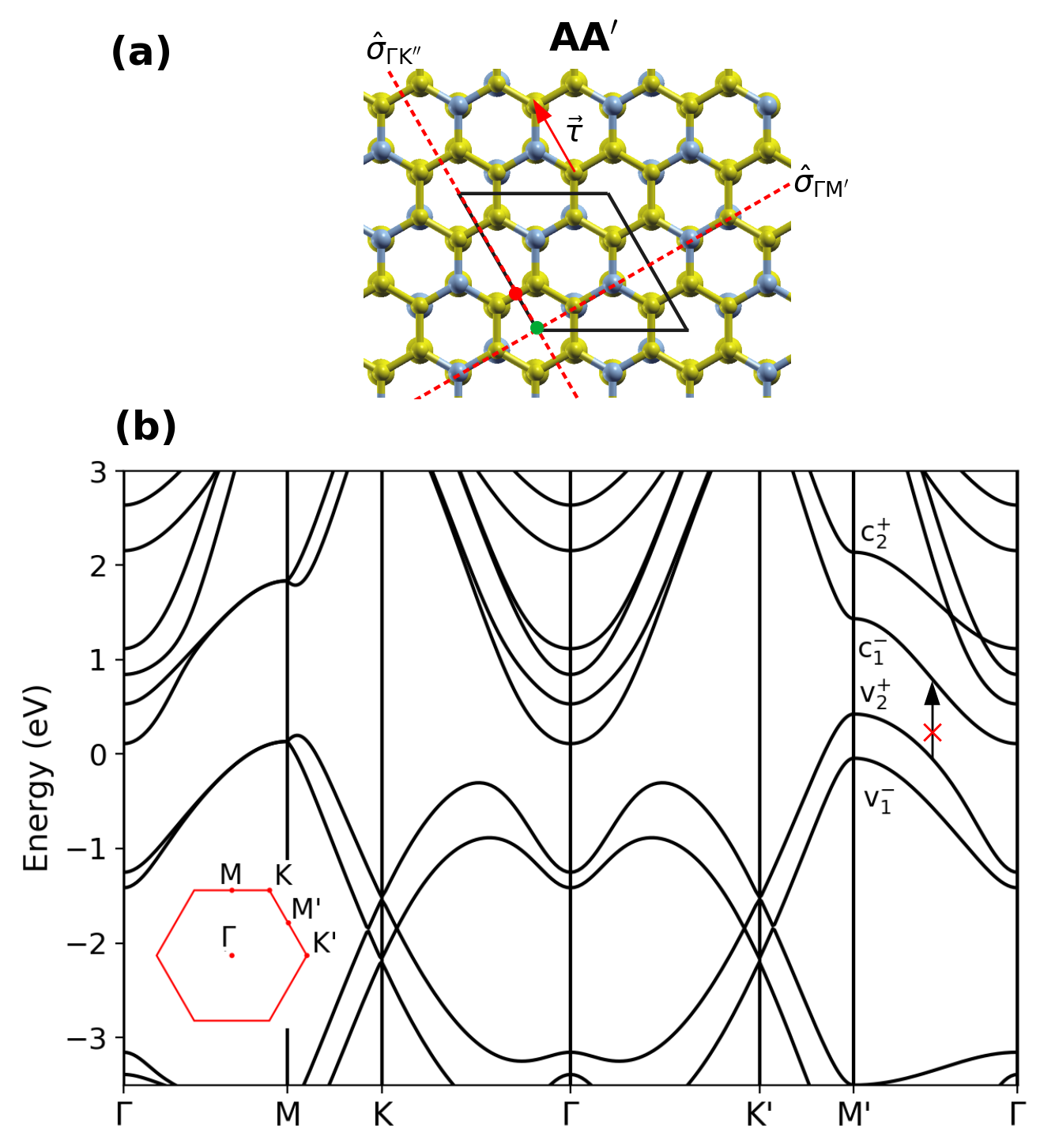}
    \caption{Crystal structure for bilayer C$_3$N with AA$'$ stacking is shown in (a). The red dot indicates the in-plane position of the inversion symmetry center, the dashed red lines represent mirror symmetry planes parallel to the stacking direction and the green dot denotes the in-plane position of the two-fold rotation axis parallel to the stacking direction. Finally, the red arrow represents the fractional translation $\Vec{\tau}$ discussed in the main test. The DFT-PBE electronic structure is shown in (b) along the direction $\Gamma\mathrm{M}'$ the highest valence and the lowest conduction bands are labelled according to their parity w.r.t. $\{\sigma_{xy}|\boldsymbol{\tau}\}$ symmetry operation.}
    \label{fig:Fig_struct_BL_AA1}
\end{figure}

%
%
As discussed in the Introduction, together with AB and AB$'$ stackings, another stable bilayer C$_3$N motif is AA$'$. Previous DFT calculations have effectively shown that these three stackings exhibit similar energies and coexistence of these motifs is expected at room temperature\cite{Wu_2018}. 
The crystal structure of AA$'$-C$_3$N is shown in Fig.~\ref{fig:Fig_struct_BL_AA1}(a). This stacking has an inversion symmetry center (shown by the red dot in the figure), two mirror symmetry planes, represented by red dashed lines, and a two-fold rotation axis parallel to the stacking direction (indicated by the green dot).
Interestingly, this stacking-motif is also invariant w.r.t. the non-symmorphic symmetry operation $\{\sigma_{xy}|\boldsymbol{\tau}\}$, corresponding to $z\rightarrow-z$ mirror symmetry followed by fractional translation of the vector $\boldsymbol{\tau}$, represented by the red arrow.

Structural optimization performed within PBE-D2 provides an in-plane lattice parameter $a$ = 4.849 \AA, while the interlayer distance is equal to 3.22 \AA, similarly to the other two stackings.  For this relaxed atomic structure, the electronic structure at the DFT level is shown in Fig.~\ref{fig:Fig_struct_BL_AA1}(b). We notice that the highest valence state (found at the $\mathrm{M}'$ point) has a higher energy than the lowest conduction state, occurring at the $\Gamma$ point, in agreement with the DFT results of Ref.~[\onlinecite{Wu_2018}] (our results give a negative "gap" $\Gamma_{c}$ - M$'_v$ = -0.31 eV). Such metallicity is a problem related to the use of the PBE functional within the Kohn-Sham DFT scheme, since, experimentally, AA$'$-C$_3$N has been shown to have a finite gap of about 0.4 eV~\cite{Wei_2021} and hybrid DFT calculations~\cite{Wei_2021} provide a semiconducting ground state.

The use of such metallic ground state to compute both QP corrections and optical properties is problematic, as it would induce a fictitious over-screening effect once the electron-electron screened interaction is evaluated using RPA approximation, providing inaccurate values for electronic gaps and exciton binding energies. 
Nevertheless, in the following, we will assume that the Kohn-Sham states computed at the PBE level are anyhow a good approximation for electronic wavefunctions, despite the  problems of the associated Kohn-Sham energies.

Figure \ref{fig:Fig_struct_BL_AA1}(b) indicates that the lowest direct band gap occurs almost in the middle of the $\Gamma\mathrm{M}'$ direction. As a consequence, it is reasonable to expect that the low energy transitions which contribute to the lowest energy excitons also come from this portion of the BZ. Notably, $\mathbf{k}$ points along this direction are invariant both under the $\hat{\sigma}_{\Gamma\mathrm{M}'}$ mirror symmetry and $\{\sigma_{xy}|\boldsymbol{\tau}\}$, therefore the electronic states can be properly labelled according to how they transform under these operations. Our DFT results indicate that the highest valence band is even under $\{\sigma_{xy}|\boldsymbol{\tau}\}$, while the lowest conduction is odd. In Fig.~\ref{fig:Fig_struct_BL_AA1}(b) these two states are indicated as $v^{+}_2$ and $c^{-}_1$, respectively. As the dipole operator $\mathbf{d}_{\epsilon}$ is invariant under $\{\sigma_{xy}|\boldsymbol{\tau}\}$ (assuming, as usual, incoming light with polarization direction $\hat{\epsilon}$ orthogonal to the stacking direction $z$), the matrix element $\langle\varphi_{v_2\mathbf{k}}|\hat{\epsilon}\cdot\mathbf{r}|\varphi_{c_1\mathbf{k}}\rangle$ is zero, independently of the direction of the polarization versor $\hat{\epsilon}$. Therefore, we expect light-induced scattering between these bands to be forbidden by symmetry and consequently the low-energy excitons composed by these transitions to be optically dark.

To confirm this symmetry-based analysis, we solve BSE computing the static electron-electron interaction in the direct kernel using PBE single particle wavefunctions and applying a rigid scissor $s^{W}_0$ to all the unoccupied bands, to manually remove DFT-spurious metallicity. Such scissor parameter has been chosen so that the minimum gap $\Gamma_{c}$ - M$'_v$ of the resulting band-structure was positive but smaller than the one found experimentally, to avoid under-screening effects. 
Furthermore, in the independent-particle part of the excitonic Hamiltonian, Eq.~\eqref{eq:Exc_H}, we mimic quasi-particle corrections via a scissor operator applied to the DFT bands, manually chosen to obtain a minimum indirect band gap $\Gamma_{c}$ - M$'_v$ equal to the experimental one (0.4 eV). We underline that such scissor is therefore larger than the one introduced in the calculation of electronic screening.

\begin{figure}
  \centering
    \includegraphics[width=0.5\textwidth]{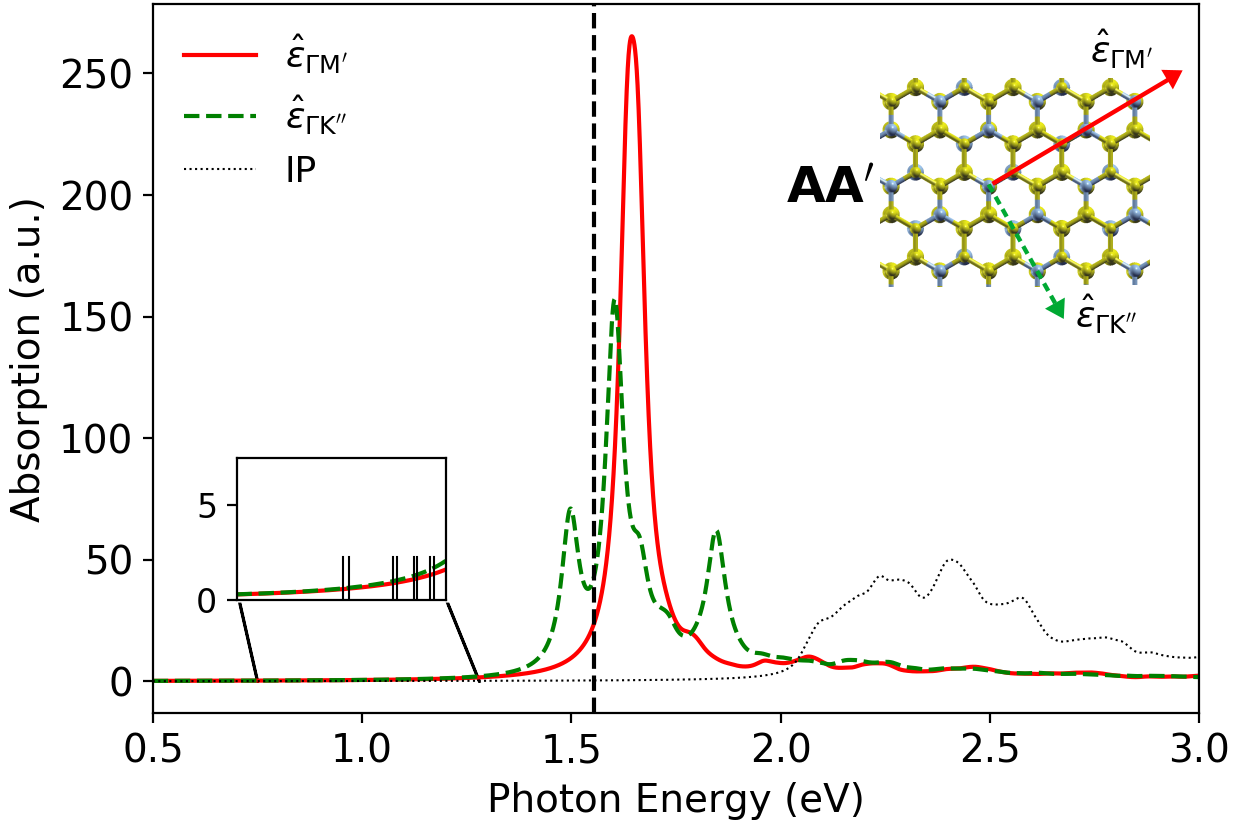}
    \caption{Absorption spectra for AA$'$-C$_3$N, computed using rigid shift parameter $s^{W}_0$ equal to 0.35 eV. Polarization directions are shown on top of the crystal structure, while the vertical dashed line indicates the energy of the minimum direct gap. Finally, the black dotted line represents the independent particle absorption spectra. The inset underlines the dark nature of low-lying excitons, whose spectral positions are represented by vertical black bars.}
    \label{fig:figura_Assorbimento_AA1}
\end{figure}

The results of these calculations are shown in Fig.~\ref{fig:figura_Assorbimento_AA1}, where we have chosen $s^{W}_0 = 0.35$ eV
, which fulfills the above-mentioned requirement to induce a small, positive band gap. The obtained spectra confirm the symmetry-based discussion just outlined. 
In fact, in both cases, low energy excitons (whose positions are indicated by black vertical bars in the inset) are optically dark independently of the polarization direction, and no absorption structure is observed in the low-energy region between 0.75 eV and 1.25 eV, where low-lying discrete excitons are found. 
The dark bound excitons are due to single particle transitions between bands $v^{+}_2$ and $c^{-}_1$ along the $\Gamma$-M$'$ direction. This point is further clarified in the \suppinfo, where, for completeness, we report the exciton wavefunctions of the low-lying excitons in reciprocal space.
We point out that the observed optical quenching is stable with respect to small variations of the $s^{W}_0$ scissor parameter. In the \suppinfo{} this is further confirmed by solving BSE using a RPA screening obtained with $s^{W}_0 = $0.45 eV.

We emphasize that a more complete analysis of the optical properties of AA$'$ stacking would require a better starting point for G$_0$W$_0$ and BSE calculations (i.e. a semiconducting electronic ground state). This is presently beyond the scope of this work and is left for future investigations.

%
%
%
\section{Conclusions}
In this work we discuss electronic and optical properties of bilayer C$_3$N using state-of-the-art \textit{ab initio} calculations. As a first point, we find that optical absorption around 1.7 eV is favoured for all stacking structures studied. Furthermore, our results suggest a particular behaviour of BL-C$_3$N,
i.e. the absence of low energy absorption peaks due to strongly bound excitons composed by electron-hole transitions with energies close to the minimum electronic direct gap. 

These findings are explained in terms of independent particle effects as due to the negligible interband dipole between the lowest conduction and the topmost valence bands involved in the formation of the involved excitons. In the case of AB and AB$'$ stackings, we develop a model for the single particle states of interest, to demonstrate that the overall interband dipole assumes negligible values because of the destructive interference of the contributions coming from the two layers. Furthermore, we also justify the quenching of low-energy absorption in AA$'$-C$_3$N stacking, exploiting the symmetry properties of the crystal. 

This work paves the way to future theoretical and experimental investigations on multilayer C$_3$N. On the one hand, it could be fascinating to investigate how optical properties of C$_3$N can be tuned varying the number of stacked monolayers or changing the twist angle between them. On the other hand, the abundant presence of dark or quasi-dark low-energy excitons in BL-C$_3$N could have important effects on exciton lifetimes and dynamics.
%
\section*{ACKNOWLEDGEMENTS}
%
M.Z. thanks Marco Gibertini and Alberto Guandalini for stimulating discussions during this work and Nicola Spallanzani and Claudia Cardoso for support in the compilation of \textit{ab initio} codes on different clusters.
Lattice structures have been displayed using the {\tt XCrySDen}~\cite{KOKALJ_1999} program. 
This work was partially funded by 
MaX -- MAterials design at the eXascale -- a European Centre of Excellence funded by the European Union's program 
HORIZON-EUROHPC-JU-2021-COE-01 (Grant No. 101093374). It was also supported by the Italian national program PRIN2017 No.~2017BZPKSZ "Excitonic insulator in two-dimensional long-range interacting systems" (EXC-INS). 

We acknowledge ICSC – Centro Nazionale di Ricerca in High Performance Computing, Big Data and Quantum Computing and ECOSISTER - Ecosystem for Sustainable Transition in Emilia-Romagna, funded under the National Recovery and Resilience Plan (NRRP) - NextGenerationEU - Mission 04, Component 2, Investments 1.4 and 1.5 [Call for tender n. 3277 dated 30/12/2021. Award Number:0001052 dated 23/06/2022].

M.J.C. acknowledges Brazilian agency CNPq and INEO (Nat. Inst. Sci. Technol. Organic Electronics - CNPq, FAPESP). Computational time on the Marconi100 machine at CINECA was provided by the Italian ISCRA program. We acknowledge EuroHPC Joint Undertaking for awarding us access to MeluXina at LuxProvide, Luxembourg. The authors gratefully acknowledge the LuxProvide teams for their expert support.
\appendix
%
%
%
%
%
\section{Properties of intralayer Hamiltonian}
\label{sec:AppendixB}
%
%
We consider the block diagonal Hamiltonian $\mathbf{h}_{\mathrm{IN}}$ defined in Section \ref{sec:INTERBAND_DIPOLE}. For each 2p$_z$ orbital on layer L$_1$ at position $\tau_{\alpha_1}$ there is an analogous state localized at $\tau_{\alpha_2} = \hat{I}\tau_{\alpha_1}$ on layer L$_2$, because of the inversion symmetry of the bilayer (with both AB and AB$'$ stacking). Thus, the diagonal blocks can be related to each other, i.e.
\begin{equation}
    \begin{split}
    H^{\mathrm{L}_1}_{\alpha_1,\beta_1}(\mathbf{k}) &= \sum_{\mathbf{R}}e^{i\mathbf{k}\cdot\mathbf{R}} \,t(\alpha_10,\beta_1\mathbf{R})\\
    &=\sum_{\mathbf{R}}e^{-i\mathbf{k}\cdot\mathbf{R}} \,t(\alpha_20,\beta_2\mathbf{R})\\
    &=H^{\mathrm{L}_2}_{\alpha_2,\beta_2}(\mathbf{k})^{*}
    \end{split}
    \label{eq:Eq_B1}
\end{equation}
where we have used the fact that the Hamiltonian of the system ${H}$ is invariant under spatial inversion symmetry and the hopping integrals are real because of time reversal and $\hat{\mathrm{I}}|\beta_1\mathbf{R}\rangle = -|\beta_2\hat{\mathrm{I}}\mathbf{R}\rangle$. Equation~\eqref{eq:Eq_B1} indicates that the matrix $\mathbf{H}^{\mathrm{L}_1}$ is the complex conjugate of $\mathbf{H}^{\mathrm{L}_2}$, therefore these matrices have the same eigenvalues. As the spectrum of $\mathbf{h}_{\mathrm{IN}}$ is the union of the spectra of $\mathbf{H}^{\mathrm{L}_1}$ and $\mathbf{H}^{\mathrm{L}_2}$, each eigenvalue $\varepsilon_{n\mathbf{k}}^0$ of $\mathbf{h}_{\mathrm{IN}}$ will be twofold degenerate, for each $\mathbf{k}$.

Furthermore, as $\mathbf{H}^{\mathrm{L}_1} = \mathbf{H}^{\mathrm{L}_2*}$, we can associate to each eigenvalue $\varepsilon_{n\mathbf{k}}^0$ the pair of eigenstates

\begin{equation}
    \begin{split}
        |\phi^{\mathrm{L}_1}_{n\mathbf{k}}\rangle &= \sum_{\alpha_1}c_{\alpha_1}(n\mathbf{k})
        |\alpha_1\mathbf{k}\rangle, \\
        |\phi^{\mathrm{L}_2}_{n\mathbf{k}}\rangle &= \sum_{\alpha_2}c_{\alpha_2}(n\mathbf{k})
        |\alpha_2\mathbf{k}\rangle,
    \end{split}
    \label{eq:Eq_B2}
\end{equation}
where 
%
$|\alpha\mathbf{k}\rangle = \frac{1}{\sqrt{N}}\sum_{\mathbf{R}}e^{i\mathbf{k}\cdot\mathbf{R}}|\alpha\mathbf{R}\rangle$ 
and $c_{\alpha_2}(n\mathbf{k}) = c_{\alpha_1}(n\mathbf{k})^*$, being
\begin{equation}
    \sum_{\beta_1}H^{\mathrm{L}_1}_{\alpha_1\beta_1}(\mathbf{k})c_{\beta_1}(n\mathbf{k}) = \varepsilon_{n\mathbf{k}}^0 c_{\alpha_1}(n\mathbf{k}).
    \label{eq:Eq_B3}
\end{equation}
We notice that $|\phi^{\mathrm{L}_1}_{n\mathbf{k}}\rangle$ ($|\phi^{\mathrm{L}_2}_{n\mathbf{k}}\rangle$) is a Bloch function localized on layer L$_1$ (L$_2$) as it only involves 2p$_z$ orbitals localized on that layer. Further, defining the time inversion operator $\hat{\mathrm{T}} = \hat{\mathrm{K}}$, i.e. equal to the complex conjugate operator, one can show that $|\phi^{\mathrm{L}_2}_{n\mathbf{k}}\rangle =- \hat{\mathrm{I}}\cdot\hat{\mathrm{T}}|\phi^{\mathrm{L}_1}_{n\mathbf{k}}\rangle$. In fact, 
\begin{equation}
    -\hat{I}\cdot\hat{T}|\phi^{L_1}_{n\mathbf{k}}\rangle = -\sum_{\alpha_1}c_{\alpha_1}(n\mathbf{k})^{*}\hat{I}\cdot\hat{T}
  |\alpha_1\mathbf{k}\rangle
    \label{eq:eq_B4}
\end{equation}
By using the reality of 2p$_z$ orbitals together with the relation: $\hat{I}|\alpha_1\mathbf{R}\rangle = -|\alpha_2\hat{I}\mathbf{R}\rangle$ we find
\begin{equation}
  \hat{I}\cdot\hat{T}|\alpha_1\mathbf{k}\rangle
    = -\frac{1}{\sqrt{N}}\sum_{\mathbf{R}}e^{i\mathbf{k}\cdot\mathbf{R}}|\alpha_2\mathbf{R}\rangle.
    \label{eq:eq_B5}
\end{equation}
Combining Eq.~\eqref{eq:eq_B4} and Eq.~\eqref{eq:eq_B5} and reminding $c_{\alpha_2}(n\mathbf{k}) = c_{\alpha_1}(n\mathbf{k})^*$ we finally obtain $|\phi^{\mathrm{L}_2}_{n\mathbf{k}}\rangle =- \hat{\mathrm{I}}\cdot\hat{\mathrm{T}}|\phi^{\mathrm{L}_1}_{n\mathbf{k}}\rangle$.

%
\renewcommand{\emph}{\textit}
\bibliographystyle{apsrev4-2}
\bibliography{bib/MAIN_BIB,bib/suppinfo} 
\end{document}